\documentclass[showpacs,aps,superscriptfootnote,12pt]{revtex4}
\usepackage{bbm}%
\usepackage{amsfonts}%
\usepackage{amsmath}%
\usepackage{graphicx}%
\usepackage{amssymb}%
\usepackage{multirow}

\begin{document}
\title{Understanding the charmed states recently observed by the LHCb and BaBar Collaborations in the quark model}

\author{Qi-Fang L\"{u} and De-Min Li}\email{lidm@zzu.edu.cn}
\affiliation{Department of Physics, Zhengzhou University, Zhengzhou,
Henan 450001, China}

\begin{abstract}

Comparing the expected spectroscopy in the relativistic quark model
and the predicted strong decays in the $^3P_0$ model employing the
realistic wave functions from the relativistic quark model with the
measured properties of the LHCb and BaBar charmed states, we find
that the masses and strong decays of these charmed states can be
reasonably explained in the conventional $q\bar{q}$ picture, and
therefore suggest that the $D(2550)$/$D_J(2580)$,
$D^*(2600)$/$D^*_J(2650)$, $D(2750)$/$D_J(2740)$,
$D^*(2760)$/$D^*_J(2760)$, $D_J(3000)$, and $D^*_J(3000)$ can be
identified as the $D(2^1S_0)$, $D(2^3S_1)$, $D^\prime_2(1D)$,
$D(1^3D_3)$, $D(3^1S_0)$, and $D(1^3F_4)$, respectively.
\end{abstract}
\pacs{12.39.Ki, 14.40.Lb, 12.38.Lg, 13.25.Ft} \maketitle

\section{Introduction}{\label{introduction}}

In 2013, the LHCb Collaboration reported several $D_J$ resonances
by studying the $D\pi$ and $D^*\pi$ final states in
$pp$ collisions\cite{Lhcb}. The $D_J(2580)$, $D^*_J(2650)$,
$D_J(2740)$, and $D_J(3000)$ were observed in the
$D^{*+}\pi^-$ channel, the $D^*_J(3000)$ was observed in the $D^+\pi^-$ channel, and
the $D^*(2760)$ was observed in both the
$D^{*+}\pi^-$ and $D^+\pi^-$ channels. The helicity-angle distributions
indicate that the $D_J(2580)$, $D_J(2740)$, and $D_J(3000)$ are the
unnatural parity resonances [$P=(-1)^{(J+1)}$], while the $D^*_J(2650)$ and
$D^*_J(2760)$ are the natural parity resonances [$P=(-1)^J$]. The observation of
the $D^*_J(3000)$ in the $D^+\pi^-$ channel makes this state should
be a natural parity resonance.

In 2010, the BaBar Collaboration also reported several charmed
states by analyzing the $D\pi$ and $D^*\pi$ systems in inclusive
$e^+e^-\rightarrow c\bar{c}$ interactions\cite{babar}. The $D(2550)$ and
$D(2750)$  were observed in the $D^{*+}\pi^-$ channel, the $D^*(2760)$
was observed in the $D^+\pi^-$ channel, and the $D^*(2600)$ was
observed in both the $D^{*+}\pi^-$ and $D^+\pi^-$ channels. The
helicity-angle distributions indicate that the
$D(2550)$ and $D(2750)$ are the unnatural parity resonances, while
the $D^*(2600)$ is the natural parity resonance. The observation of
the $D^*(2760)$ in the $D^+\pi^-$ channel shows that it should be a
natural parity resonance. The measured masses and widths of these LHCb and BaBar
charmed states mentioned above are listed in Table \ref{exp}.
Based on the masses, decay modes, and helicity-angle distributions,
we regard that the BaBar charmed states $D(2550)$, $D^*(2600)$, $D(2750)$, and $D^*(2760)$ are in fact
compatible with the LHCb states $D_J(2580)$, $D^*_J(2650)$, $D_J(2740)$, and $D^*_J(2760)$,
respectively. The average values of the LHCb and BaBar measurements are also shown in Table~\ref{exp}.

\begin{table}[htb]
\begin{center}
\caption{ \label{exp} The neutral charge resonances observed by the
LHCb Collaboration\cite{Lhcb} and the BaBar Collaboration\cite{babar}. The N and UN stand
for the natural parity and unnatural parity, respectively. }
\footnotesize
\begin{tabular}{lcccc}
\hline\hline
 State                              &  Channel               & Parity  & Property (MeV)                                                    & Average  (MeV) \\\hline
 $D_J(2580)^0$\cite{Lhcb}           & $D^{*+}\pi^-$          &UN       & Mass: $2579.5\pm 3.4\pm 5.5$        &\multirow{2}{4.2cm}{Mass: $2559.5\pm 2.8\pm 4.4$} \\
                                    &                        &         & Width: $177.5\pm 17.8 \pm 46.0$     &      \\
 $D(2550)^0$\cite{babar}            & $D^{*+}\pi^-$          &UN       & Mass: $2539.4\pm 4.5\pm 6.8$        &\multirow{2}{4.2cm}{Width: $153.8\pm 10.7\pm 23.9$}\\
                                    &                        &         & Width: $130\pm 12\pm 13$            &     \\\hline
 $D^*_J(2650)^0$\cite{Lhcb}         & $D^{*+}\pi^-$          &N        & Mass: $2649.2\pm 3.5\pm 3.5$        &\multirow{2}{4.2cm}{Mass: $2628.9\pm 2.1\pm 2.1$}\\
                                    &                        &         & Width: $140.2\pm 17.1\pm 18.6$      &    \\
 $D^*(2600)^0$\cite{babar}          & $D^{*+}\pi^-$, $D^+\pi^-$ & N    & Mass: $2608.7\pm 2.4\pm 2.5$        & \multirow{2}{4.2cm}{Width: $116.6\pm 9.1\pm 11.3$}\\
                                    &                           &      & Width: $93\pm 6\pm 13$              &\\\hline
 $D_J(2740)^0$\cite{Lhcb}           & $D^{*+}\pi^-$             & UN   & Mass: $2737.0\pm 3.5\pm 11.2$       &\multirow{2}{4.2cm}{Mass: $2744.7\pm 1.9\pm 5.8$}     \\
                                    &                           &      & Width: $73.2\pm 13.4\pm25.0$        &\\
 $D(2750)^0$\cite{babar}            & $D^{*+}\pi^-$             & UN   & Mass: $2752.4\pm 1.7\pm 2.7$        &\multirow{2}{4.2cm}{Width: $72.1\pm 7.4\pm 13.7$ }\\
                                    &                           &      & Width: $71\pm 6\pm 11$              &\\\hline
 $D^*_J(2760)^0$\cite{Lhcb}        & $D^{*+}\pi^-$, $D^+\pi^-$  & N    & Mass: $2761.1\pm 5.1\pm 6.5$        &\multirow{2}{4.2cm}{Mass: $2762.2\pm 2.8\pm 3.5$}     \\
                                   &                            &      & Width: $74.4\pm 3.4\pm 37.0$        &\\
 $D^*(2760)^0$\cite{babar}         & $D^+\pi^-$                 & N    & Mass: $2763.3\pm 2.3\pm 2.3$        &\multirow{2}{4.2cm}{Width: $67.6\pm 3.1\pm 18.6$}\\
                                   &                            &      & Width: $60.9\pm 5.1\pm 3.6$          &\\\hline
 $D_J(3000)^0$\cite{Lhcb}          & $D^{*+}\pi^-$               & UN      & Mass: $2971.8\pm 8.7$   & Mass: $2971.8\pm 8.7$ \\
                                   &                            &           & Width: $188.1 \pm 44.8$&Width: $188.1 \pm 44.8$\\\hline
 $D^*_J(3000)^0$\cite{Lhcb}        & $D^+\pi^-$                  & N       & Mass: $3008.1\pm 4.0$   &Mass: $3008.1\pm 4.0$\\
                                   &                             &         & Width: $110.5 \pm 11.5$ & Width: $110.5 \pm 11.5$\\\hline\hline
\end{tabular}
\end{center}
\end{table}

Apart from the ordinary $q\bar{q}$ states, other exotic states such
as glueballs, hybrids, and multiquark systems are expected to exist
in the framework of Quantum Chromodynamics (QCD). The identification
of these exotic states requires to understand well the conventional
$q\bar{q}$ meson spectroscopy both theoretically and
experimentally. To a large extent, our knowledge of meson
spectroscopy is based on some phenomenological QCD motivated models
such as quark models which are widely accepted to offer the most
complete description of meson properties and are probably the most
successful phenomenological models of hadron
structures\cite{godfreyRMP}.

According to the PDG\cite{Beringer:1900zz}, the low-mass charmed
mesons $D(1^1S_0)$, $D(1^3S_1)$, $D(1^3P_1)$, $D(1^1P_1)$,
$D(1^3P_0)$, and $D(1^3P_2)$ predicted by quark models are well
established experimentally, however, many other higher radial and
orbital excitations of $D$ mesons predicted by quark models have not
yet been established. These charmed states recently reported by the
LHCb Collaboration and the BaBar Collaboration are clearly of
importance to improve our understanding of the charmed meson
spectroscopy. The possible $q\bar{q}$ quark-model assignments for
these observed charmed states have been studied in the context of
various models such as the chiral quark
model\cite{Zhong:2010vq,zhong2,zhong3}, the heavy meson effective
theory\cite{Wang:2010ydc,fazio2012,Wang:2013tka}, and the $^3P_0$
model with the simple harmonic oscillator wave
functions\cite{Sun:2010pg,Li:2010vx,Sun:2013qca,Yu:2014dda,Close:2005se}
or the nonrelativistic quark model wave
functions\cite{Segovia:2012cd}, and other
approaches\cite{othercharm1,othercharm2,othercharm3}. It is also
natural and necessary to exhaust the possible $q\bar{q}$
descriptions before restoring to more exotic assignments. The
theoretical predictions for these states are not completely
consistent with the measured properties and there is not yet a
consensus on the assignments of these states. Therefore, in order to
deeply understand these newly reported charmed states, further test
calculations against the experimental measurements are still
required.

In this work, we shall compare the observed properties of the
LHCb and BaBar charmed states with the mass predictions of the relativistic quark model
and strong decay predictions of the
$^3P_0$ model employing the relativistic quark model wave functions to
determine their spectroscopic assignments.

This work is organized as follows. In Sec.~II, we calculate the
charmed meson masses in the Godfrey and Isgur (GI) relativized quark model and give
the possible assignments for these charmed states based on their
observed masses and decay modes. In Sec.~III, we investigate the
strong decays of these states for different possible assignments in
the $^3P_0$ model using the realistic wave functions from the GI quark model.
 The summary is given in the last section.

\section{Masses}{\label{masses}}

To understand the properties of the charmed mesons, we shall discuss their masses
in a relativistic quark model
proposed by Godfrey and Isgur\cite{Godfrey:1985xj}. In this model,
the total Hamiltonian is
\begin{equation}
\tilde{H} = H_0+\tilde{V}(\boldsymbol{p},\boldsymbol{r}),\label{GI}
\end{equation}
\begin{equation}
H_0 = (\boldsymbol{p}^2+m_1^2)^{1/2}+(\boldsymbol{p}^2+m_2^2)^{1/2},
\end{equation}
\begin{equation}
\tilde{V}(\boldsymbol{p},\boldsymbol{r}) =
\tilde{H}^{\text{conf}}_{12}+\tilde{H}^{\text{cont}}_{12}+\tilde{H}^{\text{ten}}_{12}+\tilde{H}^{\text{so}}_{12},
\end{equation}
where $\tilde{H}^{\text{conf}}_{12}$ includes the spin-independent
linear confinement and Coulomb-type interactions;
$\tilde{H}^{\text{cont}}_{12}$,  $\tilde{H}^{\text{ten}}_{12}$, and $\tilde{H}^{\text{so}}_{12}$ are the
 color contact term, color tensor interaction, and spin-orbit interaction, respectively. The details of this model
and the explicit form of these interactions
can be found in Appendix A of Ref.\cite{Godfrey:1985xj}.

The spin-orbit interaction term $\tilde{H}^{\text{so}}_{12}$ can decompose into
symmetric $\tilde{H}^{\text{so}}_{(12)}$ and antisymmetric $\tilde{H}^{\text{so}}_{[12]}$.
The antisymmetric $\tilde{H}^{\text{so}}_{[12]}$ can cause the
the mixing of the charmed mesons with different total spins but with the same total angular momentum such as
$D(n{}^3L_L)$ and $D(n{}^1L_L)$.  Consequently, the two physical states $D_L(nL)$ and $D^\prime_L(nL)$ can be
described by\cite{Godfrey:1985xj,Godfrey:1986wj}
\begin{equation}
\left(
\begin{array}{cr}
D_L(nL)\\
D^\prime_L(nL)
\end{array}
\right)
 =\left(
 \begin{array}{cr}
\cos \theta_{nL} & \sin \theta_{nL} \\
-\sin \theta_{nL} & \cos \theta_{nL}
\end{array}
\right)
\left(\begin{array}{cr}
D(n^1L_L)\\
D(n^3L_L)
\end{array}
\right),
\label{ddmixing1}
\end{equation}
where the $\theta_{nL}$ is the mixing angle.

The total Hamiltonian $\tilde{H}$ can be divided into the diagonal
part $\tilde{H}_{\text{diag}}=
H_0+\tilde{H}^{\text{conf}}_{12}+\tilde{H}^{\text{cont}}_{12}+(\tilde{H}^{\text{ten}}_{12})_{\text{diag}}
+\tilde{H}^{\text{so}}_{(12)}$ and the off-diagonal part
$\tilde{H}_{\text{off}}=\tilde{H}^{\text{so}}_{[12]}+(\tilde{H}^{\text{ten}}_{12})_{\text{off}}$, where
$(\tilde{H}^{\text{ten}}_{12})_{\text{off}}$ can cause $^3L_J\leftrightarrow$ $^3L^\prime_J$ mixing and
is neglected in the present work.  We use the
Gaussian expansion method \cite{Hiyama:2003cu} to solve the
Hamiltonian (\ref{GI}). To actually perform the calculations, we
first diagonalize $\tilde{H}_{\text{diag}}$ in the Gaussian function
basis to obtain the masses and wave functions for the unmixed $D$ mesons,
then in the basis $|n^{2S+1}L_J\rangle$, we diagonalize the off-diagonal part
$\tilde{H}^{\text{so}}_{[12]}$, which is treated as the perturbative
term, to obtain the masses of the mixed $D_L$ and $D^\prime_L$ mesons.
In this procedure, the mixing angle $\theta_{nL}$ can be obtained.
The values of the model parameters used in our calculations are taken
from Ref.\cite{Godfrey:1985xj}.

We apply the GI quark model to calculate the mass spectra of the
$1S$, $2S$, $3S$, $1P$, $2P$, $3P$, $1D$, $2D$, $3D$, $1F$, $2F$,
$3F$, $1G$, and $2G$ states. The masses of the $1S$, $2S$, $1P$,
$1D$, and $1F$ states in the origin paper of Godfrey and
Isgur\cite{Godfrey:1985xj} are well reproduced. Our calculated $D$
meson masses are listed in Table \ref{tmass}. The predictions of
some other relativistic quark models\cite{zvr,de,efg} are also
listed. We do not list the mass predictions of
Ref.\cite{Sun:2013qca}, where the masses of the $D_L(nL)$ and
$D^\prime_L(nL)$ are not calculated due to the
$\tilde{H}^{\text{so}}_{[12]}$ term being neglected and the unmixed
$D$ meson spectra are almost the same as our present calculations.
These predictions from different relativistic quark models give us a
mass range for the corresponding $D$ meson, which can restrict the
possible assignments for the reported charmed states.

\begin{table}[htbp]
\begin{center}
\caption{ \label{tmass} The $D$ meson masses in MeV from
different relativistic quark models. The mixing angles of $D_L-D^\prime_L$ obtained in
this work are $\theta_{1P}=-25.5^\circ$, $\theta_{2P}=-29.4^\circ$, $\theta_{3P}=-27.7^\circ$, $\theta_{1D}=-38.2^\circ$,
$\theta_{2D}=-37.5^\circ$, $\theta_{3D}=-36.8^\circ$, $\theta_{1F}=-39.5^\circ$, $\theta_{2F}=-39.4^\circ$,
$\theta_{3F}=-39.4^\circ$, $\theta_{1G}=-40.2^\circ$, and $\theta_{2G}=-40.3^\circ$. A dash denotes
that the corresponding mass was not calculated in the corresponding reference.}
\footnotesize
\begin{tabular}{lcccc|lcccc}
\hline\hline
  State         & This work  & ZVR\cite{zvr}    & DE\cite{de}     &EFG\cite{efg}      &State             & This work    &ZVR\cite{zvr}     &DE\cite{de}     &EFG\cite{efg} \\\hline
  $D(1^1S_0)$      & 1874       & 1850          & 1868            & 1871              & $D(2^3D_3)$      & 3227         & 3190             &$-$                & 3335    \\
  $D(1^3S_1)$      & 2038       & 2020          & 2005            & 2010              & $D(3^3D_1)$      & 3595         & $-$                &$-$                &$-$         \\
  $D(2^1S_0)$      & 2583       & 2500          & 2589            & 2581              & $D_2(3D)$        & 3576         & $-$                &$-$                &$-$         \\
  $D(2^3S_1)$      & 2645       & 2620          & 2692            & 2632              & $D^\prime_2(3D)$ & 3610         & $-$                &$-$                &$-$         \\
  $D(3^1S_0)$      & 3068       & 2980          & 3141            & 3062              & $D(3^3D_3)$      & 3591         & $-$                &$-$                &$-$         \\
  $D(3^3S_1)$      & 3111       & 3070          & 3226            & 3096              & $D(1^3F_2)$      & 3132         & 3000             & 3101            & 3090    \\
  $D(1^3P_0)$      & 2398       & 2270          & 2377            & 2406              & $D_3(1F)$        & 3109         & 3010             & 3074            & 3129    \\
  $D_1(1P)$        & 2455       & 2400          & 2417            & 2426              & $D^\prime_3(1F)$ & 3144         & 3030             & 3123            & 3145    \\
  $D^\prime_1(1P)$ & 2467       & 2410          & 2490            & 2469              & $D(1^3F_4)$      & 3113         & 3030             & 3091            & 3187    \\
  $D(1^3P_2)$      & 2501       & 2460          & 2460            & 2460              & $D(2^3F_2)$      & 3491         & 3380             &$-$                &$-$         \\
  $D(2^3P_0)$      & 2932       & 2780          & 2949            & 2919              & $D_3(2F)$        & 3462         & 3390             &$-$                &$-$         \\
  $D_1(2P)$        & 2925       & 2890          & 2995            & 2932              & $D^\prime_3(2F)$ & 3499         & 3410             &$-$                &$-$         \\
  $D^\prime_1(2P)$ & 2961       & 2890          & 3045            & 3021              & $D(2^3F_4)$      & 3466         & 3410             &$-$                &3610    \\
  $D(2^3P_2)$      & 2957       & 2940          & 3035            & 3012              & $D(3^3F_2)$      & 3833         & $-$                &$-$                & $-$        \\
  $D(3^3P_0)$      & 3344       & 3200          & $-$               & 3346              & $D_3(3F)$        & 3809         & $-$                &$-$                & $-$        \\
  $D_1(3P)$        & 3329       & 3290          & $-$               & 3365              & $D^\prime_3(3F)$ & 3843         & $-$                &$-$                & $-$        \\
  $D^\prime_1(3P)$ & 3362       & 3300          & $-$               & 3461              & $D(3^3F_4)$      & 3816         & $-$                &$-$                & $-$        \\
  $D(3^3P_2)$      & 3356       & 3340          & $-$               & 3407              & $D(1^3G_3)$      & 3398         & 3240             &$-$                & 3352    \\
  $D(1^3D_1)$      & 2816       & 2710          & 2795            & 2788              & $D_4(1G)$        & 3365         & 3240             &$-$                & 3403    \\
  $D_2(1D)$        & 2816       & 2740          & 2775            & 2806              & $D^\prime_4(1G)$ & 3400         & 3260             &$-$                & 3415    \\
  $D^\prime_2(1D)$ & 2845       & 2760          & 2833            & 2850              & $D(1^3G_5)$      & 3362         & $-$                &$-$                & 3473    \\
  $D(1^3D_3)$      & 2833       & 2780          & 2799            & 2863              & $D(2^3G_3)$      & 3722         & $-$                &$-$                & $-$        \\
  $D(2^3D_1)$      & 3232       & 3130          & $-$               & 3228              & $D_4(2G)$        & 3687         & $-$                &$-$                & $-$        \\
  $D_2(2D)$        & 3212       & 3160          & $-$               & 3259              & $D^\prime_4(2G)$ & 3723         & $-$                &$-$                & $-$        \\
  $D^\prime_2(2D)$ & 3249       & 3170          & $-$               & 3307              & $D(2^3G_5)$      & 3685         & $-$                &$-$                & 3860    \\\hline\hline

\end{tabular}
\end{center}
\end{table}

\begin{figure}[htbp]
\includegraphics[scale=0.5]{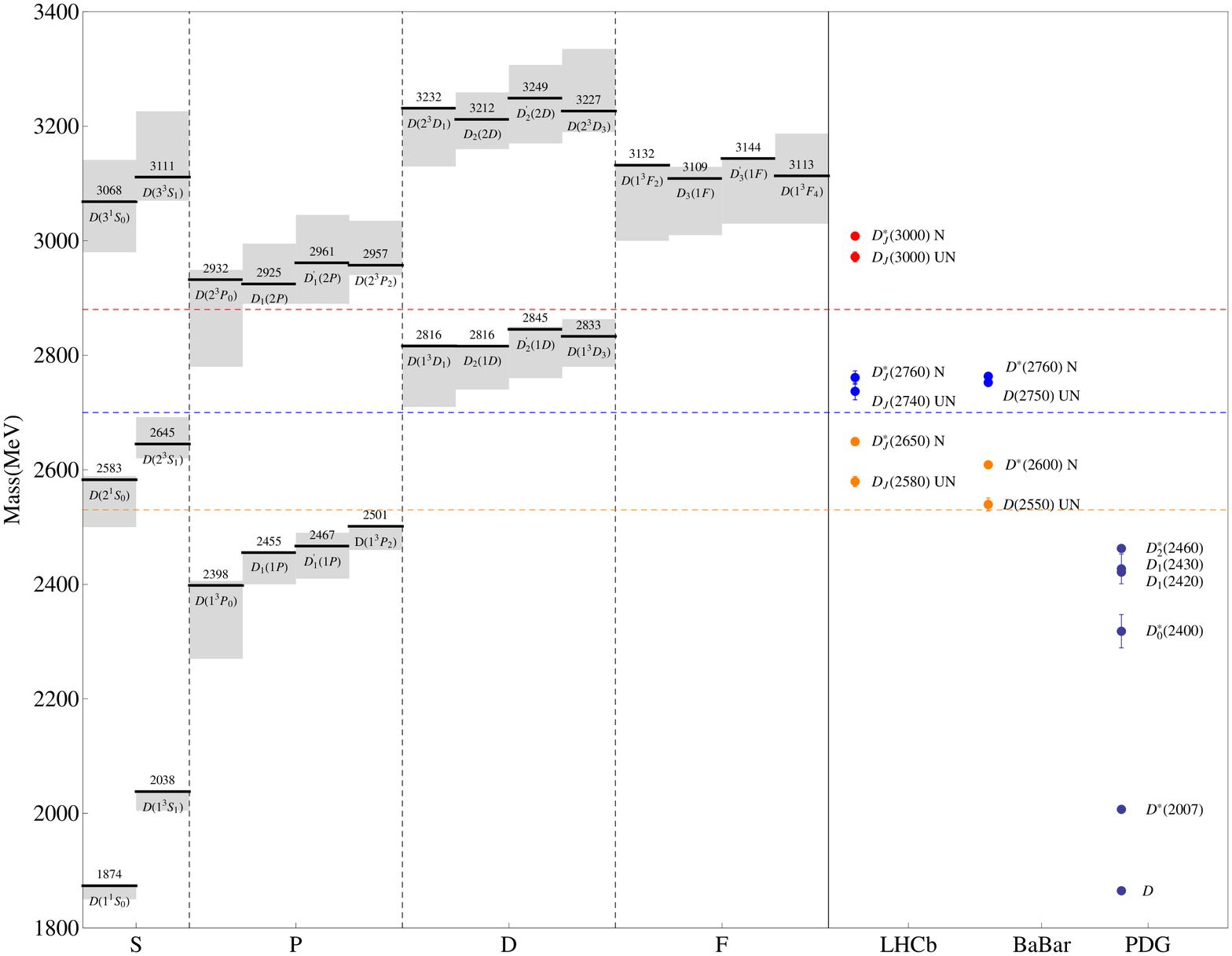}
\vspace{-0.6cm} \caption{The charmed meson spectrum. The solid lines are the GI quark model predictions and the
shaded regions are the expected mass ranges from some other relativistic quark models\cite{zvr,de,efg}.
The observed charmed states are also shown. The N and UN denote natural parity and unnatural parity, respectively.}
\label{fmass}%
\end{figure}

The predicted mass ranges from different relativistic quark models
and information of the observed charmed states are depicted in Fig.~\ref{fmass}.
Clearly, the masses of the $D$, $D^*(2007)$, $D^*_0(2400)$,
$D_1(2420)$, $D_1(2430)$, and $D^*_2(2460)$ as the well-established
ground charmed mesons are well reproduced. The $D(2550)/D_J(2580)$ and $D^*(2600)/D^*_J(2650)$ lie within
the $D(2^1S_0)$ and $D(2^3S_1)$ mass ranges, respectively. The $D(2750)/D_J(2740)$ and $D^*(2760)/D^*_J(2760)$
lie within the $1D$ states mass ranges. The $D_J(3000)$ and $D^*_J(3000)$ lie close to
 the mass ranges of the $3S$, $2P$, and $1F$ states.  We regard the BaBar state $D^*(2760)$
and the LHCb state $D^*_J(2760)$ as the same state, so the observation of the $D^*_J(2760)$ in the $D^{*+}\pi$
channel excludes the $D(2^3P_0)$ assignment, although the expected
$D(2^3P_0)$ mass range is very close to the $D^*(2760)$/$D^*_J(2760)$ mass. We don't consider the possibility
of the $D^*_J(3000)$ and $D_J(3000)$ being the
the $2D$ states, since the predicted $2D$ mass ranges are at least 130 MeV higher than the $D^*_J(3000)$
and $D_J(3000)$ masses. Based on the mass and the parity,
our tentative assignments for these newly
reported charmed states are listed in Table \ref{assignment}. Below, we shall focus on these possible assignments.
It should be noted that the mass information alone is insufficient to classify these charmed states, so their decay behaviors also need to be compared
with model expectations. In the next section, we shall discuss the strong decays in the $^3P_0$ model employing the
wave functions from the GI quark model.

\begin{table}
\begin{center}
\caption{\label{assignment} Possible assignments for the LHCb and BaBar charmed
states based on masses and decay modes} \footnotesize
\begin{tabular}{ll}
\hline\hline
 State                       &  Possible assignments  \\\hline
$D(2550)/D_J(2580)$          & $D(2^1S_0)$            \\
$D^*(2600)/D^*_J(2650)$      & $D(2^3S_1)$\\
$D(2750)/D_J(2740)$          & $D_2(1D)$, $D^\prime_2(1D)$\\
$D^*(2760)/D^*_J(2760)$      & $D(1{}^3D_1)$, $D(1{}^3D_3)$\\
$D_J(3000)$                  & $D(3^1S_0)$, $D_1(2P)$, $D^\prime_1(2P)$, $D_3(1F)$, $D^\prime_3(1F)$\\
$D^*_J(3000)$                & $D(3^3S_1)$, $D(2^3P_0)$, $D(2^3P_2)$, $D(1^3F_2)$, $D(1^3F_4)$\\
\hline\hline
\end{tabular}
\end{center}
\end{table}

\section{Strong decays}

\subsection{$^3P_0$ model}

In this section, we employ the $^3P_0$ model to evaluate the
Okubo-Zweig-Iizuka-allowed two-body strong decays of the
initial state. The $^3P_0$ model, also known as the quark pair creation model, in which meson decay takes place
through a quark-antiquark pair with the vacuum quantum number\cite{micu}, has been extensively
applied to study the strong decay of
hadrons. There exists exhaustive literature on the $^3P_0$ model and some detailed reviews on the $^3P_0$ model
can be found in Refs.\cite{3p0model1,3p0model2,3p0model3,3p0model4}. Here we give the
main ingredients of the $^3P_0$ model. Following the conventions in Ref.\cite{3p0decay}, the transition operator $T$ of the decay $A\rightarrow BC$ in the
$^3P_0$ model is given by
\begin{eqnarray}
T=-3\gamma\sum_m\langle 1m1-m|00\rangle\int
d^3\boldsymbol{p}_3d^3\boldsymbol{p}_4\delta^3(\boldsymbol{p}_3+\boldsymbol{p}_4){\cal{Y}}^m_1
\left(\frac{\boldsymbol{p}_3-\boldsymbol{p}_4}{2}\right
)\chi^{34}_{1-m}\phi^{34}_0\omega^{34}_0b^\dagger_3(\boldsymbol{p}_3)d^\dagger_4(\boldsymbol{p}_4),
\end{eqnarray}
where $\gamma$ is a dimensionless $q_3\bar{q}_4$ pair-production strength, and $\boldsymbol{p}_3$ and
$\boldsymbol{p}_4$ are the momenta of the created quark $q_3$ and
antiquark  $\bar{q}_4$, respectively. $\phi^{34}_{0}$,
$\omega^{34}_0$, and $\chi_{{1,-m}}^{34}$ are the flavor, color,
and spin wave functions of the  $q_3\bar{q}_4$, respectively. The
solid harmonic polynomial
${\cal{Y}}^m_1(\boldsymbol{p})\equiv|p|^1Y^m_1(\theta_p, \phi_p)$ reflects
the momentum-space distribution of the $q_3\bar{q}_4$ .

The partial wave amplitude ${\cal{M}}^{LS}(\boldsymbol{P})$ for $A\rightarrow BC$ is expressed as
\begin{eqnarray}
{\cal{M}}^{LS}(\boldsymbol{P})&=&
\sum_{\renewcommand{\arraystretch}{.5}\begin{array}[t]{l}
\scriptstyle M_{J_B},M_{J_C},\\\scriptstyle M_S,M_L
\end{array}}\renewcommand{\arraystretch}{1}\!\!
\langle LM_LSM_S|J_AM_{J_A}\rangle\langle
J_BM_{J_B}J_CM_{J_C}|SM_S\rangle\nonumber\\
&&\times\int
d\Omega\,\mbox{}Y^\ast_{LM_L}{\cal{M}}^{M_{J_A}M_{J_B}M_{J_C}}
(\boldsymbol{P}), \label{pwave}
\end{eqnarray}
where ${\cal{M}}^{M_{J_A}M_{J_B}M_{J_C}}
(\boldsymbol{P})$ is the helicity amplitude and defined by
\begin{eqnarray}
\langle
BC|T|A\rangle=\delta^3(\boldsymbol{P}_A-\boldsymbol{P}_B-\boldsymbol{P}_C){\cal{M}}^{M_{J_A}M_{J_B}M_{J_C}}(\boldsymbol{P}).
\end{eqnarray}
The $|A\rangle$, $|B\rangle$, and $|C\rangle$ stand for the mock meson states.  The mock meson $|A\rangle$
is defined by\cite{mockmeson}
\begin{eqnarray}
|A(n^{2S_A+1}_AL_{A}\,\mbox{}_{J_A M_{J_A}})(\boldsymbol{P}_A)\rangle
&\equiv& \sqrt{2E_A}\sum_{M_{L_A},M_{S_A}}\langle L_A M_{L_A} S_A
M_{S_A}|J_A
M_{J_A}\rangle\nonumber\\
&&\times  \int d^3\boldsymbol{p}_A\psi_{n_AL_AM_{L_A}}(\boldsymbol{p}_A)\chi^{12}_{S_AM_{S_A}}
\phi^{12}_A\omega^{12}_A\nonumber\\
&&\times  \left|q_1\left({\scriptstyle
\frac{m_1}{m_1+m_2}}\boldsymbol{P}_A+\boldsymbol{p}_A\right)\bar{q}_2
\left({\scriptstyle\frac{m_2}{m_1+m_2}}\boldsymbol{P}_A-\boldsymbol{p}_A\right)\right\rangle,
\end{eqnarray}
where $m_1$ and $m_2$ ($\boldsymbol{p}_1$ and $\boldsymbol{p}_2$) are the masses (momenta) of the
quark $q_1$ and the antiquark $\bar{q}_2$, respectively; $\boldsymbol{P}_A=\boldsymbol{p}_1+\boldsymbol{p}_2$,
$\boldsymbol{p}_A=\frac{m_2\boldsymbol{p}_1-m_1\boldsymbol{p}_2}{m_1+m_2}$;
$\chi^{12}_{S_AM_{S_A}}$, $\phi^{12}_A$, $\omega^{12}_A$,
$\psi_{n_AL_AM_{L_A}}(\boldsymbol{p}_A)$ are the spin, flavor, color, and
space wave functions of the meson $A$ composed of $q_1\bar{q}_2$ with total energy $E_A$, respectively.

Various $^3P_0$ models exist in literature and typically differ in the
choice of the pair-production vertex, the phase space conventions, and the
meson wave functions employed. In this work, we restrict to the simplest vertex as introduced originally by
Micu\cite{micu} which assumes a
spatially constant pair-production strength $\gamma$,
adopt the relativistic phase space as Ref.\cite{3p0decay}, and employ the realistic meson wave functions
from the GI quark model.
With the relativistic phase space, the decay width
$\Gamma(A\rightarrow BC)$ in terms of the partial wave
amplitude Eq.~(\ref{pwave}) is given by
\begin{eqnarray}
\Gamma(A\rightarrow BC)= \frac{\pi
P}{4M^2_A}\sum_{LS}|{\cal{M}}^{LS}(\boldsymbol{P})|^2, \label{width1}
\end{eqnarray}
where $P=|\boldsymbol{P}|$=$\sqrt{[M^2_A-(M_B+M_C)^2][M^2_A-(M_B-M_C)^2]}/2M_A$,
and $M_A$, $M_B$, and $M_C$ are the masses of the meson $A$, $B$,
and $C$, respectively.

In order to determine the phase space and final state momenta, the
masses of the initial state mesons involved in this work are taken
to be the average values of Table \ref{exp}, and the masses of the
final state mesons, except for the theoretical candidates of the
final state mesons such as $D(2^1S_0)$, $D(2^3S_1)$, $D(1^3D_1)$,
$D(1^3D_3)$, $D(1D)$, and $D^\prime(1D)$, are taken from the
PDG\cite{Beringer:1900zz}. These theoretical candidates masses are:
$M_{D(2^1S_0)}=(M_{D(2550)}+M_{D_J(2580)})/2$,
$M_{D(2^3S_1)}=(M_{D^*(2600)}+M_{D^*_J(2650)})/2$,
$M_{D(1^3D_1)}\simeq
M_{D(1^3D_3)}=(M_{D^*(2760)}+M_{D^*_J(2760)})/2$, $M_{D_2(1D)}\simeq
M_{D^\prime_2(1D)}= (M_{D_J(2740)}+M_{D(2750)})/2$. The mixing
angles $\theta_{nL}$ are taken from Table \ref{tmass}, and the
mixing angle of $D_{s1}(2460)$-$D_{s1}(2536)$ is solved to be
$-37.5^\circ$. There is only one free parameter $\gamma$ in our
calculations. We set
  $\gamma=8.9$ by fitting to the following 34 two-body decay modes with
  specific branching ratios\cite{Beringer:1900zz}: (1) $a_2(1320)\rightarrow \eta\pi$, (2) $a_2(1320)\rightarrow
K\bar{K}$, (3) $f_2(1270)\rightarrow\pi\pi$, (4)
$f_2(1270)\rightarrow K\bar{K}$, (5) $f^\prime_2(1525)\rightarrow
K\bar{K}$, (6) $f^\prime_2(1525)\rightarrow \eta\eta$, (7)
$\pi_2(1670)\rightarrow f_2(1270)\pi$, (8) $\pi_2(1670)\rightarrow
\rho\pi$, (9) $\pi_2(1670)\rightarrow KK^*(892)+c.c.$, (10)
$\pi_2(1670)\rightarrow\omega\rho$,
  (11) $\rho_3(1690)\rightarrow\omega\pi$,
  (12) $\rho_3(1690)\rightarrow\pi\pi$, (13) $\rho_3(1690)\rightarrow
  K\bar{K}$, (14) $f_4(2050)\rightarrow\pi\pi$, (15) $K^*(1410)\rightarrow K\pi$, (16) $K^*_0(1430)\rightarrow
K\pi$, (17) $K^*_2(1430)\rightarrow K\pi$, (18)
$K^*_2(1430)\rightarrow K^*(892)\pi$, (19) $K^*_2(1430)\rightarrow
K\rho$, (20) $K^*_2(1430)\rightarrow K\omega$, (21)
$K^*(1680)\rightarrow K\pi$, (22) $K^*(1680)\rightarrow K\rho$, (23)
$K^*(1680)\rightarrow K^*(892)\pi$, (24) $K^*_3(1780)\rightarrow
K\rho$, (25) $K^*_3(1780)\rightarrow K^*(892)\pi$, (26)
$K^*_3(1780)\rightarrow K\pi$, (27) $K^*_3(1780)\rightarrow K\eta$,
(28) $K^*_4(2045)\rightarrow K\pi$, (29) $K^*_4(2045)\rightarrow
\phi K^*(892)$, (30) $D^*(2010)^+\rightarrow D^0\pi^+$, (31)
$D^*(2010)^+\rightarrow D^+\pi^0$, (32) $\psi(3770)\rightarrow
D\bar{D}$, (33) $\Upsilon(4S)\rightarrow B^+B^-$, and (34)
$\Upsilon(4S)\rightarrow B^0\bar{B}^0$. Here $\gamma$ denotes the
  light nonstrange quark pair $u\bar{u}$ or $d\bar{d}$
  creation strength and the strange quark pair $s\bar{s}$ creation
  strength $\gamma_{s\bar{s}}$ can be related by $\gamma_{s\bar{s}}=\gamma\frac{m_u}{m_s}$\cite{rss}, where $m_u$ and $m_s$
  are respectively the
  $u$-quark and $s$-quark masses employed in the GI quark model. Our value of $\gamma$ is higher than
  that used by other groups such as\cite{Close:2005se, 3p0model4} by a factor of $\sqrt{96\pi}$ due to different field
  conventions, constant factor in the transition operator $T$, etc.

\subsection{$D(2550)$/$D_J(2580)$}

The decay widths of the $D(2550)$/$D_J(2580)$ as the $D(2^1S_0)$
compared with the experiment data are shown in Table~\ref{D2550}.
Under the $D(2^1S_0)$ assignment, the $D(2550)$/$D_J(2580)$ is predicted to have a total width of 128 MeV,  which
is in good agreement with $153.8\pm 10.7\pm 23.9$ MeV, the average value of the LHCb and BaBar measurements.
The dominant decay channel of the $D(2^1S_0)$ is expected to be
the $D^{*+}\pi^-$, which naturally explains that the $D(2550)$ and $D_J(2580)$ were observed in
the $D^{*+}\pi^-$ channel. Therefore, both the masses and decays support that the
 $D(2550)$/$D_J(2580)$ is in fact the resonance $D(2^1S_0)$.

\begin{table}
\begin{center}
\caption{ \label{D2550} Decay widths of the $D(2550)$/$D_J(2580)$ as the $D(2^1S_0)$ in MeV.}
\scriptsize
\begin{tabular}{lc}
\hline\hline
  $D^{*+}\pi^-$                                 & 85.20\\
  $D^{*0}\pi^0$                                 & 43.22\\
  $D_0^*(2400)^0\pi^0$                          & 0.05 \\
  $D_0^*(2400)^+\pi^-$                          & 0.09 \\
  $D^*\eta$                                     & 0.12 \\
  Total width                                   & 128.69 \\
  Experiment  & $153.8\pm 10.7\pm 23.9$  \\
\hline\hline
\end{tabular}
\end{center}
\end{table}

\subsection{$D^*(2600)$/$D^*_J(2650)$}

In Table \ref{D2600}, we list the decay widths of the $D^*(2600)$/$D^*_J(2650)$
as the $D(2^3S_1)$ state.  The predicted total width of the $D(2^3S_1)$
is 122 MeV, in good agreement with the average value of the LHCb and BaBar widths, $116.6\pm 9.1\pm 11.3$ MeV.
 The predicted branching ratio
\begin{eqnarray}
\frac{\Gamma(D^*(2600)\rightarrow D^+\pi^-)}{\Gamma(D^*(2600)\rightarrow D^{*+}\pi^-)}=0.43
\end{eqnarray}
is consistent with the BaBar result of $0.32\pm0.02\pm0.09$\cite{babar}. Also,
the expected dominant decay modes of the $D(2^3S_1)$ are $D^*\pi$ and $D\pi$,
which is consistent with the observation. So, the masses and
decays support
that the $D^*(2600)$/$D^*_J(2650)$ is the state $D(2^3S_1)$.

\begin{table}
\begin{center}
\caption{ \label{D2600} Decay widths of the $D^*(2600)$/$D^*_J(2650)$ as the $D(2^3S_1)$ in MeV.}
\scriptsize
\begin{tabular}{lcc}
\hline\hline
  $D^0\pi^0$            & 10.82              \\
  $D^+\pi^-$            & 21.66              \\
  $D_sK$                & 3.30               \\
  $D\eta$               & 4.73               \\
  $D^{*0}\pi^0$         & 25.20               \\
  $D^{*+}\pi^-$         & 49.82               \\
  $D^*\eta$             & 3.71                \\
  $D_s^*K$              & 0.55                \\
  $D_2^*(2460)^0 \pi^0$ & 4.8$\times10^{-3}$  \\
  $D_2^*(2460)^+ \pi^-$ & 5.6$\times10^{-3}$  \\
  $D_1(2430)^0 \pi^0$   & 0.59               \\
  $D_1(2430)^+ \pi^-$   & 1.16               \\
  $D_1(2420)^0 \pi^0$   & 0.27               \\
  $D_1(2420)^+ \pi^-$   & 0.50               \\
    Total width           & 122.30              \\
  Experiment            & $116.6\pm9.1\pm11.3$
 \\
\hline\hline
\end{tabular}
\end{center}
\end{table}

\subsection{$D_J(2740)$/$D(2750)$}

The decay widths of the $D_J(2740)$/$D(2750)$ as the $D_2(1D)$ and
$D^\prime_2(1D)$ are listed in Table \ref{tD2740}. The total width
for the $D_2(1D)$ with a mass around 2750 MeV is expected to be
about 280 MeV, much larger than the LHCb and BaBar results, which
excludes the assignment of the $D_J(2740)$/$D(2750)$ as the
$D_2(1D)$ state.
 With the $D^\prime_2(1D)$ assignment to the $D_J(2740)$/$D(2750)$, its total width
is predicted to be 109 MeV, which is reasonably close to the LHCb
and BaBar experimental average of $72.1\pm 7.4\pm 13.7$ MeV, and
consistent with the LHCb width of $73.2\pm 13.4\pm 25.0$
MeV\cite{Lhcb}. In other approaches such as the chiral quark
model\cite{zhong2,zhong3} and the heavy quark effective
theory\cite{othercharm2}, it is also found that under the
$D^\prime_2(1D)$ assignment, the $D_J(2740)$/$D(2750)$ width can be
reasonably accounted for.

It is noted that the mass prediction of the GI quark model for the
$D^\prime_2(1D)$ differs by around $100$ MeV from the observed mass
of the $D_J(2740)$/$D(2750)$. This discrepancy between the GI quark
model and the experiment maybe result from that the coupled channel
effects (also called hadron loop effects) were neglected in the GI
quark model. The coupled channel effects can give rise to mass
shifts to the bare hadron states. The mass shifts induced by the
coupled channel effects can present a better description of the $D$,
$D_s$, charmonium, and bottomonium states\cite{othercharm1,loops}.
Other approaches such as the nonrelativistic quark
model\cite{Li:2010vx}, the relativistic quark model\cite{zvr}, and
the Blankenbecler-Sugar equation\cite{lnr} consistently predict that
the $D^\prime_2(1D)$ mass is very close to the $D_J(2740)/D(2750)$
mass, which make the $D^\prime_2(1D)$ assignment to the
$D_J(2740)$/$D(2750)$ possible based on its mass.

Therefore, identification of the $D_J(2740)$/$D(2750)$ as the
$D^\prime_2(1D)$ state seems favored by the available experimental
data. The conclusion that the $D_J(2740)$/$D(2750)$ can be explained
as the $D^\prime_2(1D)$ has been suggested by
Refs.\cite{zhong2,zhong3,Wang:2010ydc,fazio2012,Wang:2013tka,Li:2010vx,Yu:2014dda,othercharm2}.

In the heavy quark effective theory, the $D_2(1D)$ and
$D^\prime_2(1D)$ can be grouped into the $(1^-,2^-)_{j=\frac{3}{2}}$
and $(2^-,3^-)_{j=\frac{5}{2}}$ doublets, where $j$ is the total
angular momentum of the light quark. In the heavy quark limit, there
are two mixing angle\cite{jjcoupling}: one is $-50.8^\circ$ for
which the $D^\prime_2(1D)$ belongs to the $j=\frac{5}{2}$ states,
and the other is $39.2^\circ$ for which the $D_2(1D)$ belongs to the
$j=\frac{3}{2}$ states. Among these two $2^-$ charmed mesons, the
decay width is expected to be broader for $j=\frac{3}{2}$ than for
$j=\frac{5}{2}$.

\begin{figure}[htbp]
\includegraphics[scale=0.8]{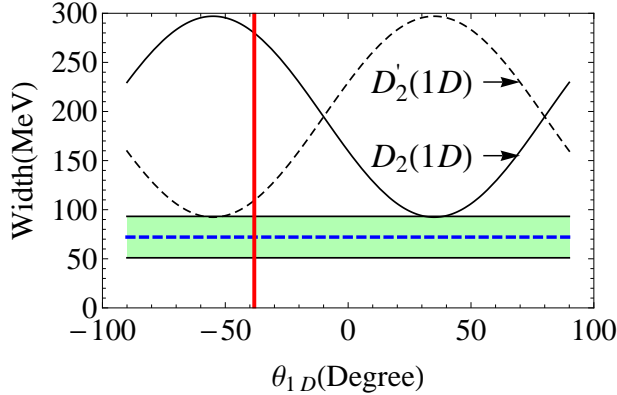}
\vspace{-0.6cm} \caption{Total decay width of $D_J(2740)$/$D(2750)$ as
the $D_2(1D)$ and $D^\prime_2(1D)$ versus the mixing angle. The blue dashed line with a
green band denotes the LHCb and BaBar experimental average. The vertical red solid line corresponds to
the mixing angle $\theta_{1D}=-38.2^\circ$ obtained in Sec.~\ref{masses}.}
\label{fD2740}
\end{figure}

The total widths
of the $D_J(2740)$/$D(2750)$ as the $D_2(1D)$ and $D^\prime_2(1D)$ dependence on the
mixing angle $\theta_{1D}$ are also depicted in
Fig. \ref{fD2740}. It is clear that for the $\theta_{1D}$ lying around
$(-20^\circ\sim-80^\circ)$, the $D_2(1D)$ is much broader
than the $D^\prime_2(1D)$. Our predicted $\theta_{1D}=-38.2^\circ$ is
close to $-50.8^\circ$ while far from $39.2^\circ$, so, with the $D^\prime_2(1D)$ assignment,
$D_J(2740)/D(2750)$ corresponds to the $2^-$ charmed meson belonging to the $(2^-,3^-)_{j=\frac{5}{2}}$ doublet.

The dominant decay channels of the $D^\prime_2(1D)$ are expected to be $D^*\pi$, $D\rho$,
and $D\omega$, while the dominant
decay channels of the $D_2(1D)$ are $D^*\pi$, $D^*\eta$,
$D^*_2\pi$. The $D^*_2\pi$ channel is the most dominant decay mode of the $D_2(1D)$, and therefore
is the ideal decay channel to further experimental search for the partner of the $D_J(2740)/D(2750)$.

\begin{table}
\begin{center}
\caption{ \label{tD2740} Decay widths of the $D_J(2740)$/$D(2750)$ as the $D_2(1D)$ and $D^\prime_2(1D)$ in MeV.}
\scriptsize
\begin{tabular}{lcc}
\hline
\hline
 Channel                        & $D_2(1D)$     & $D^\prime_2(1D)$ \\
\hline
  $D^{*0}\pi^0$                  & 38.21       & 7.00   \\
  $D^{*+}\pi^-$                  & 75.98       & 13.50  \\
  $D^*\eta$                      & 13.34       & 0.93  \\
  $D_s^*K$                       & 6.40        & 0.35  \\
  $D_2^*(2460)^0 \pi^0$          & 48.57       & 2.36   \\
  $D_2^*(2460)^+ \pi^-$          & 96.17       & 4.66  \\
  $D_1(2430)^0 \pi^0$            & 0.03        & 0.02   \\
  $D_1(2430)^+ \pi^-$            & 0.05        & 0.04   \\
  $D_1(2420)^0 \pi^0$            & 0.10        & 0.31   \\
  $D_1(2420)^+ \pi^-$            & 0.19        & 0.57  \\
  $D^0 \rho^0$                   & 0.23        & 20.78  \\
  $D^+ \rho^-$                   & 0.39        & 39.39   \\
  $D \omega$                     & 0.18        & 19.34  \\
  $D^*_0(2400)^0\pi^0$           & 0.02        & 0.11  \\
  $D^*_0(2400)^+\pi^-$           & 0.02        & 0.07  \\
  Total width                    & 279.90      & 109.43   \\
  Experiment            &\multicolumn{2}{c}{$72.1\pm7.4\pm13.7$}  \\
\hline
\hline
\end{tabular}
\end{center}
\end{table}

\subsection{$D^*(2760)$/$D^*_J(2760)$}

The decay widths of the $D^*(2760)$/$D^*_J(2760)$ as the $D(1^3D_1)$
and $D(1^3D_3)$ are listed in Table \ref{tD2760}. Our results show
the possibility of the $D^*(2760)$/$D^*_J(2760)$ being the
$D(1^3D_1)$ can be excluded because the theoretical total width is
366 MeV, much larger than the LHCb and BaBar results. With the
$D(1^3D_3)$ assignment, the $D^*(2760)$/$D^*_J(2760)$ is expected to
have a width of about 28 MeV, which is somewhat lower than the
average value of the LHCb and BaBar results but roughly consistent
with the LHCb width of $74.4\pm 3.4\pm 37.0$ MeV\cite{Lhcb}. The
discrepancy between the predicted width of the $^3P_0$ model for the
$D(1^3D_3)$ and the average width of the $D^*(2760)$ and
$D^*_J(2760)$ could arise from the inherent uncertainty of the
$^3P_0$ model itself. As pointed out by Ref.\cite{3p0model3}, the
$^3P_0$ model is a coarse model of the complicated strong decay
theory and the best it can hope for is to predict a decay width to
within a factor of 2. Apart from the $^3P_0$ model, other approaches
such as the chiral quark model\cite{zhong2,zhong3} and the heavy
quark effective theory\cite{othercharm2} have been used to discuss
the possibility of the $D^*(2760)$/$D^*_J(2760)$ being the
$D(1^3D_3)$. In Refs.\cite{zhong2,zhong3}, the
$D^*(2760)$/$D^*_J(2760)$ as the $D(1^3D_3)$ is expected to have a
width of about 68 MeV, consistent with the experiment. In
Ref.\cite{othercharm2}, the $D^*(2760)$/$D^*_J(2760)$ as the
$D(1^3D_3)$ is expected to have a width of about 40 MeV, very close
to the lower limit of $67.6\pm 3.1\pm 18.6$ MeV, the average value
of the LHCb and BaBar results. So, the measured widths in fact
strongly prefer the $D(1^3D_3)$ over the $D(1^3D_1)$ assignment to
the $D^*(2760)$/$D^*_J(2760)$.

Similar to the case of the $D_J(2740)/D(2750)$, the observed
$D^*(2760)$/$D^*_J(2760)$ mass is about 70 MeV lower than the GI
quark model mass prediction for the $D(1^3D_3)$. As mentioned in
subsection D, the coupled channel effects being neglected in the GI
quark model maybe cause the discrepancy between the experiment and
the GI quark model. Other model calculations such as nonrelativistic
quark model\cite{Li:2010vx}, coupled channel
effects\cite{othercharm1}, the relativistic quark model\cite{zvr},
and Regge phenomenology\cite{liregge} show that the $D(1^3D_3)$ mass
is close to the $D^*(2760)$/$D^*_J(2760)$ mass, which makes the
$D(1^3D_3)$ assignment to the $D^*(2760)$/$D^*_J(2760)$ plausible
based on its mass.

With the assignment of the $D(2750)$ as the $D^\prime_2(1D)$, the predicted width ratio is
\begin{eqnarray}
\frac{\Gamma(D^*(2760)\rightarrow D^+\pi^-)}{\Gamma(D(2750)\rightarrow D^{*+}\pi^-)}=\left\{\begin{array}{c}
0.68, D^*(2760)=D(1^3D_3)\\
4.09, D^*(2760)=D(1^3D_1)
\end{array}\right..
\end{eqnarray}
Comparison of the theoretical ratio and the corresponding BaBar
experimental ratio of $0.42\pm 0.05\pm0.11$\cite{babar} also
strongly prefers the $D(1^3D_3)$ over the $D(1^3D_1)$ assignment to
the $D^*(2760)$. The suggestion that the $D^*(2760)$/$D^*_J(2760)$
can be identified as the $D(1^3D_3)$ has been proposed in
Refs.\cite{zhong2,zhong3,Wang:2010ydc,fazio2012,Wang:2013tka,Sun:2010pg,Li:2010vx,Yu:2014dda,othercharm2}
based on its mass and width.

The decay behavior of the $D(1^3D_1)$ is remarkably different from those of the $D(1^3D_3)$.
The main decay modes of the $D(1^3D_1)$ are expected to be $D\pi$, $D^*\pi$, $D_1(2420)\pi$,
$D_1(2430)\pi$, $D\rho$, $D\eta$, $D_sK$, and $D\omega$, while the $D(1^3D_3)$ is expected to mainly decay to
$D\pi$ and $D^*\pi$, consistent with the observation of the $D^*_J(2760)$ in both the $D\pi$ and $D^*\pi$ channels.
Further experimental analysis of the $D_1(2420)\pi$, $D_1(2430)\pi$, $D\rho$, $D\eta$, $D_sK$, and $D\omega$ systems
would be helpful to search for the candidate for the $D(1^3D_1)$.

\begin{table}
\begin{center}
\caption{ \label{tD2760} Decay widths of the $D^*(2760)$/$D^*_J(2760)$ as the
$D(1^3D_1)$ and $D(1^3D_3)$ in MeV.}
\scriptsize
\begin{tabular}{lcccc}
\hline
\hline

  Channel                & $D(1^3D_1)$ & $D(1^3D_3)$\\\hline
  $D^0\pi^0$             & 27.61       & 4.77                \\
  $D^+\pi^-$             & 55.24       & 9.29                \\
  $D_sK$                 & 9.47        & 0.22                \\
  $D\eta$                & 13.69       & 0.77                \\
  $D^{*0}\pi^0$          & 13.56       & 3.81                \\
  $D^{*+}\pi^-$          & 26.98       & 7.28                 \\
  $D^*\eta$              & 4.95        & 0.26                \\
  $D_s^*K$               & 2.54        & 0.04               \\
  $D_2^*(2460)^0 \pi^0$  & 0.23        & 0.22                \\
  $D_2^*(2460)^+ \pi^-$  & 0.43        & 0.40                \\
  $D_1(2430)^0 \pi^0$    & 13.76       & 0.13                \\
  $D_1(2430)^+ \pi^-$    & 27.54       & 0.24                \\
  $D_1(2420)^0 \pi^0$    & 46.92       & 0.01   \\
  $D_1(2420)^+ \pi^-$    & 93.43       & 0.02                \\
  $D^0 \rho^0$           & 7.71        & 0.24                \\
  $D^+ \rho^-$           & 14.75       & 0.42                \\
  $D \omega$             & 7.28        & 0.20                \\
  $D(2^1S_0)^0\pi^0$     & 0.17        & $2.1\times 10^{-4}$ \\
  $D(2^1S_0)^+\pi^-$     & 0.31        & $3.5\times 10^{-4}$\\
  Total width            & 366.58      & 28.32               \\
  Experiment            & \multicolumn{2}{c}{$67.6\pm3.1\pm18.6$}\\
\hline
\hline
\end{tabular}
\end{center}
\end{table}

\subsection{$D_J(3000)$}

The decay widths of the $D_J(3000)$ as the $D(3^1S_0)$, $D_1(2P)$, $D_1(2P)$, $D_3(1F)$, and $D^\prime_3(1F)$
are listed in Table \ref{tD3000}. It is clear that the most favorable assignment of the $D_J(3000)$ is the $D(3^1S_0)$,
since the predicted $D(3^1S_0)$ total width of 180 MeV agrees quite well with the experimental data
of $188.1\pm 44.8$ MeV while the total widths
for other assignments are far from the measurement. The main decay modes of the $D(3^1S_0)$ are expected to be the
$D^*_2\pi$, $D^*\pi$, $D^*\rho$, $D^*\omega$, and $D(2^3S_1)\pi$.

Our most favorable assignment of the $D_J(3000)$ is the $D(3^1S_0)$, which is inconsistent with
the $D_1(2P)$ assignment proposed by Refs.\cite{Sun:2013qca, Yu:2014dda} and the $D^\prime_3(1F)$ assignment proposed
by Ref.\cite{Yu:2014dda}. The partial widths of the
$D(2^3S_1)\pi$, $D(1^3D_1)\pi$, $D(1^3D_3)\pi$, $D_2(1D)\pi$, and
$D^\prime_2(1D)\pi$ are considered in our calculations but neglected in Refs.\cite{Sun:2013qca, Yu:2014dda}.
 Our results show that the total contributions of these
channels are large for the
$D(3^1S_0)$, $D_1(2P)$, $D^\prime_1(2P)$, and $D_3(1F)$ but tiny for the $D^\prime_3(1F)$.  Without doubt,
further experimental
studies of the
$D_J(3000)$ on the $D_1(2420)\pi$, $D_1(2430)\pi$, $D_2(1D)\pi$, and $D^\prime_2(1D)\pi$ channels will
be useful to check our present assignment because
these decay modes are forbidden for the $D(3^1S_0)$ while allowable for the $D_1(2P)$, $D^\prime_1(2P)$, $D_3(1F)$,
and $D^\prime_3(1F)$.

\begin{figure}
\includegraphics[scale=0.7]{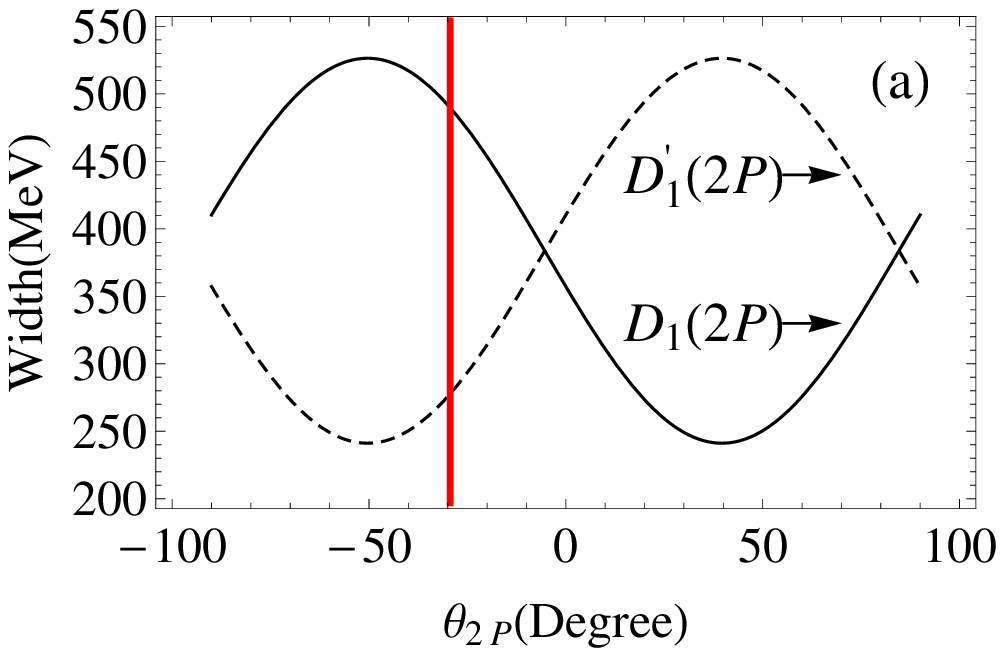}
\includegraphics[scale=0.7]{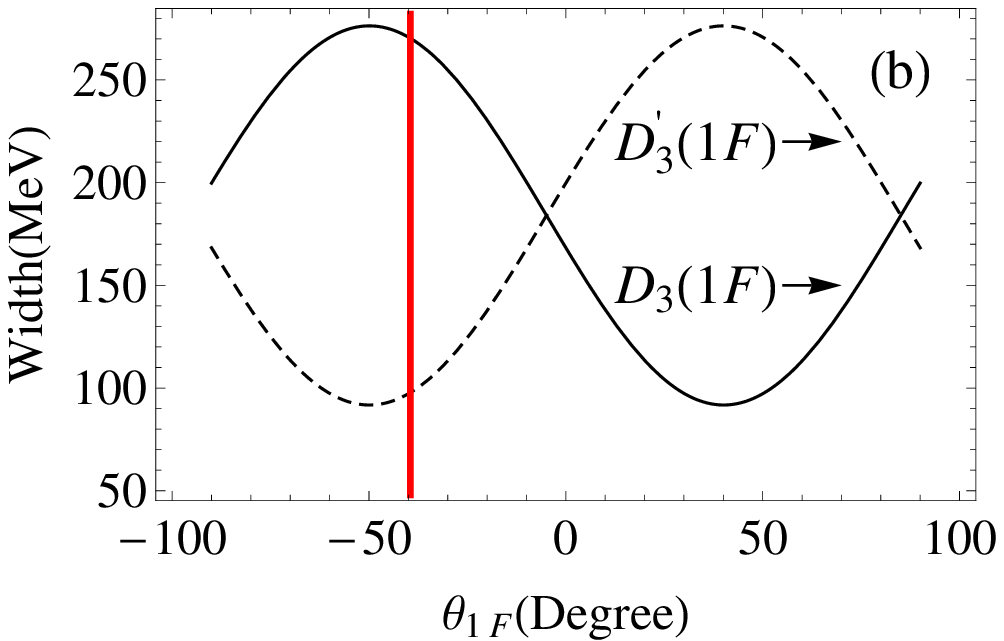}
\vspace{-0.6cm}
\caption{Total widths of the $D_1(2P)$, $D^\prime_1(2P)$, $D_3(1F)$, and $D^\prime_3(1F)$
versus the mixing angle. The vertical red solid line corresponds
to the mixing angle obtained in Sec.~\ref{masses}: $\theta_{2P}=-29.4^\circ$ and $\theta_{1F}=-39.5^\circ$.}
\label{fD3000}
\end{figure}

The total widths
of the $D_1(2P)$, $D^\prime_1(2P)$, $D_3(1F)$, and $D^\prime_3(1F)$ dependence on the
mixing angles are depicted in Fig.~\ref{fD3000}, which will be helpful to determine
the mixing angles $\theta_{2P}$ and $\theta_{1F}$ based on the measured widths.

\begin{table}
\begin{center}
\caption{ \label{tD3000} Decay widths of $D_J(3000)$ with several possible assignments in MeV.
A dash indicates that a decay mode is forbidden.}
\scriptsize
\begin{tabular}{lccccccc}\hline\hline
                        &$D(3^1S_0)$          & $D_1(2P)$            & $D^\prime_1(2P)$      & $D_3(1F)$           & $D^\prime_3(1F)$\\\hline
  $D^0 \rho^0$          & 0.13                & 26.99                & 1.16                  & 1.38                & 13.40      \\
  $D^+ \rho^-$          & 0.36                & 53.14                & 1.98                  & 2.60                & 26.34      \\
  $D \omega$            & 0.20                & 26.55                & 0.95                  & 1.28                & 13.15      \\
  $D^*_0(2400)^0\pi^0$  & 1.67                & 1.93                 & 0.94                  & 0.12                & 0.29      \\
  $D^*_0(2400)^+\pi^-$  & 4.91                & 4.06                 & 1.98                  & 0.13                & 0.29        \\
  $D^*_0(2400)\eta$     & 3.56                & 0.84                 & 0.40                  & 3.5$\times10^{-3}$ & 6.4$\times10^{-3}$    \\
  $D_2^*(2460)^0 \pi^0$ & 13.77               & 40.40                & 6.98                  & 33.26               & 1.04          \\
  $D_2^*(2460)^+ \pi^-$ & 27.07               & 80.53                & 13.70                 & 66.04               & 2.07         \\
  $D^{*0}\pi^0$         & 8.72                & 18.79                & 15.64                 & 16.18               & 4.76         \\
  $D^{*+}\pi^-$         & 16.67               & 36.92                & 31.25                 & 32.18               & 9.20          \\
  $D^*\eta$             & 0.55                & 4.39                 & 6.88                  & 6.59                & 0.83          \\
  $D^*\eta^{\prime}$    & 0.07                & 3.80                 & 0.95                  & 2.0$\times10^{-3}$ & 5.4$\times10^{-5}$  \\
  $D^{*0} \rho^0$       & 15.53               & 29.47                & 32.84                 & 5.87                & 5.46          \\
  $D^{*+} \rho^-$       & 32.13               & 57.33                & 62.68                 & 11.07               & 10.30         \\
  $D^* \omega$          & 16.18               & 28.70                & 31.31                 & 5.53                & 5.14         \\
  $D_s^*K$              & 5.7$\times10^{-3}$  & 0.95                 & 4.14                  & 2.88                & 0.20           \\
  $D_sK^*$              & 1.41                & 1.48                 & 10.41                 & 3.4$\times10^{-3}$ & 0.48           \\
  $D_{s0}^*(2317)K$     & 4.09                & 1.19                 & 0.74                  & 6.7$\times10^{-3}$ & 0.02           \\
  $D_1(2430)^0 \pi^0$   & $-$                 & 0.11                 & 0.15                  & 0.02                & 4.6$\times10^{-3}$  \\
  $D_1(2430)^+ \pi^-$   & $-$                 & 0.21                 & 0.29                  & 0.04                & 9.1$\times10^{-3}$  \\
  $D_1(2420)^0 \pi^0$   & $-$                 & 2.77                 & 11.32                 & 0.29                & 0.79              \\
  $D_1(2420)^+ \pi^-$   & $-$                 & 5.53                 & 22.62                 & 0.56                & 1.52               \\
  $D_1(2420) \eta$      & $-$                 & 7.2$\times10^{-3}$   & 0.03                  & 4.0$\times10^{-8}$ & 9.9$\times10^{-8}$    \\
  $D_{s1}(2460)K$       & $-$                 & 2.1$\times10^{-4}$   & 0.01                  & 1.3$\times10^{-8}$ & 2.7$\times10^{-6}$    \\

  $D(2^3S_1)^0\pi^0$    &11.03                &20.90                 &5.93                   &0.56                 &0.02                      \\
  $D(2^3S_1)^+\pi^-$    &21.86                &42.04                 &11.84                  &1.10                 &0.04                     \\
  $D(1^3D_1)^0\pi^0$    &0.16                 &0.02                  &0.02                   &$6.7\times 10^{-5}$  &$1.2\times 10^{-5}$                      \\
  $D(1^3D_1)^0\pi^-$    &0.29                 &0.32                  &0.04                   &$1.2\times 10^{-4}$  &$2.3\times 10^{-5}$                      \\
  $D(1^3D_3)^0\pi^0$    &$8.1\times 10^{-2}$  &0.23                  &0.01                   &27.77                &0.77                      \\
  $D(1^3D_3)^+\pi^-$    &0.01                 &0.41                  &0.02                   &54.80                &1.53                      \\
  $D_2(1D)^0\pi^0$        &$-$                  &$4.9\times 10^{-3}$   &0.01                   &$3.9\times 10^{-3}$  &$2.1\times 10^{-3}$                      \\
  $D_2(1D)^+\pi^-$        &$-$                  &$9.1\times 10^{-3}$   &0.02                   &$7.1\times 10^{-3}$  &$3.9\times 10^{-3}$                      \\
$D^\prime_2(1D)^0\pi^0$   &$-$                  &0.03                  &0.14                   &0.01                 &0.03                       \\
$D^\prime_2(1D)^+\pi^-$   &$-$                  &0.06                  &0.25                   &0.02                 &0.05                        \\

  Total width           & 180.39              & 489.85               & 277.62                &270.33               & 97.75 \\
  Experiment            & \multicolumn{5}{c}{$188.1\pm44.8$} \\\hline\hline
\end{tabular}
\end{center}
\end{table}

\subsection{$D^*_J(3000)$}

The decay widths of the $D^*_J(3000)$ as the $D(3^3S_1)$, $D(2^3P_0)$, $D(2^3P_2)$,
$D(1^3F_2)$, and $D(1^3F_4)$ are listed in Table \ref{tDx3000}. The predicted total widths
indicate that the assignments of the $D^*_J(3000)$ as the $D(2^3P_0)$, $D(2^3P_2)$,
and $D(1^3F_2)$ can be ruled out because the corresponding widths are much larger than the LHCb measurement
of $110.5\pm11.5$ MeV\cite{Lhcb}. Among the remaining two possible assignments of the $D^*_J(3000)$, the measured
width prefers the $D(1^3F_4)$ over the $D(3^3S_1)$, since the $D(3^3S_1)$ width is 157 MeV, somewhat larger than
the experimental data, while the $D(1^3F_4)$ width is 103 MeV, in good agreement with the experimental data.
The main decay modes of the $D(1^3F_4)$ are predicted to include the $D\pi$,
 $D^*\pi$, $D^*\rho$, $D^*\omega$.

 Our most favorable assignment of the $D^*_J(3000)$ is the $D(1^3F_4)$. Ref.\cite{Sun:2013qca} suggests
 that the $D^*_J(3000)$ can be explained as the $D(2^3P_0)$ and Ref.\cite{Yu:2014dda} assigns the $D^*_J(3000)$ as
 the $D(1^3F_2)$ or $D(1^3F_4)$. Obviously, the theoretical interpretations on the $D^*_J(3000)$ are not completely
 consistent with each other. The partial widths of
 the $D(2^1S_0)\pi$, $D(2^3S_1)\pi$, $D(1^3D_1)\pi$, $D(1^3D_1)\pi$, $D_2(1D)\pi$, and
$D^\prime_2(1D)\pi$ are considered in our present calculations while neglected in Refs.\cite{Sun:2013qca,Yu:2014dda}.
Our results indicate that total contributions of these channels are large for
the $D(3^1S_0)$, $D(2^3P_0)$, $D(2^3P_2)$, and $D(1^3F_2)$
while tiny for the $D(1^3F_4)$. Further experimental information of the $D^*_J(3000)$ in these
channels will be helpful to test our present assignment.

\begin{table}
\begin{center}
\caption{ \label{tDx3000} Decay widths of the $D^*_J(3000)$ with several possible assignments in MeV. A dash
indicates that a decay mode is forbidden.}
\scriptsize
\begin{tabular}{lccccc}
\hline\hline
                        &$D(3^3S_1)$         & $D(2^3P_0)$   & $D(2^3P_2)$    & $D(1^3F_2)$        & $D(1^3F_4)$\\\hline
  $D^0\pi^0$            & 4.72               & 27.98         & 0.62           & 9.52               & 3.37 \\
  $D^+\pi^-$            & 9.29               & 55.48         & 1.30           & 19.03              & 6.59   \\
  $D\eta$               & 1.39               & 9.75          & 0.93           & 4.72               &  0.83   \\
  $D^0 \rho^0$          & 0.17               & $-$           & 12.79          & 5.94               & 1.25   \\
  $D^+ \rho^-$          & 0.25               & $-$           & 25.40          & 11.71              & 2.40   \\
  $D \omega$            & 0.12               & $-$           & 12.72          & 5.85               & 1.19   \\
  $D\eta^{\prime}$      & 6.4$\times10^{-3}$ & 0.18          & 0.84           & 1.53               & 0.03   \\
  $D_2^*(2460)^0 \pi^0$ & 6.87               & $-$           & 6.23           & 4.40               & 0.86  \\
  $D_2^*(2460)^+ \pi^-$ & 13.60              & $-$           & 12.36          & 8.73               & 1.66   \\
  $D^{*0}\pi^0$         & 6.59               & $-$           & 3.91           & 7.02               & 3.21   \\
  $D^{*+}\pi^-$         & 12.76              & $-$           & 7.98           & 13.99              & 6.20   \\
  $D^*\eta$             & 0.85               & $-$           & 2.97           & 3.02               & 0.57  \\
  $D^*\eta^{\prime}$    & 0.37               & $-$           & 0.13           & 0.06               & 1.2$\times10^{-4}$ \\
  $D^{*0} \rho^0$       & 8.93               & 73.53         & 19.78          & 3.99               & 18.85  \\
  $D^{*+} \rho^-$       & 19.19              & 141.85        & 38.52          & 7.53               & 36.15 \\
  $D^* \omega$          & 9.68               & 70.98         & 19.29          & 3.76               & 18.09 \\
  $D_1(2430)^0 \pi^0$   & 1.62               & 19.46         & 4.91           & 8.11               & 0.63 \\
  $D_1(2430)^+ \pi^-$   & 3.25               & 39.01         & 9.81           & 16.23              & 1.25 \\
  $D_1(2430) \eta$      & 1.79               & 2.01          & 0.43           & 0.53               & 4.8$\times10^{-4}$ \\
  $D_1(2420)^0 \pi^0$   & 4.32               & 39.21         & 3.25           & 26.52              & 0.04 \\
  $D_1(2420)^+ \pi^-$   & 8.58               & 78.38         & 6.33           & 52.82              &  0.08 \\
  $D_1(2420) \eta$      & 0.62               & 3.86          & 0.03           & 2.15               & 1.5$\times10^{-6}$ \\
  $D_sK$                & 0.48               & 2.52          & 1.09           & 2.61               & 0.21 \\
  $D_s^*K$              & 0.08               & $-$           & 2.22           & 1.41               & 0.12 \\
  $D_sK^*$              & 0.74               & $-$           & 1.21           & 0.35               & 6.2$\times10^{-3}$ \\
  $D_s^*K^*$            & 0.09               & 3.78          & 6.68           & 7.6$\times10^{-5}$ & 4.7$\times10^{-4}$ \\
  $D_{s1}(2460)K$       & 2.65               & 1.77          & 0.83           & 0.26               & 1.6$\times10^{-3}$ \\
  $D(2^1S_0)^0\pi^0$    & 4.44               &21.57          &2.78            &1.23                &0.03                     \\
  $D(2^1S_0)^+\pi^-$    & 8.90               &43.56          &5.49            &2.43                &0.07                       \\
  $D(2^3S_1)^0\pi^0$    & 8.02               &$-$            &1.93            &0.32                &0.01                         \\
  $D(2^3S_1)^+\pi^-$    & 15.98              &$-$            &3.80            &0.62                &0.02                         \\
  $D(1^3D_1)^0\pi^0$    & 0.07               &$-$            &$6.0\times 10^{-4}$ &$1.9\times 10^{-3}$    &$1.7\times 10^{-6}$                         \\
  $D(1^3D_1)^0\pi^-$    & 0.12               &$-$            &$1.1\times 10^{-3}$ &$3.5\times 10^{-3}$    &$3.1\times 10^{-6}$                         \\
  $D(1^3D_3)^0\pi^0$    & 0.02               &$-$            &0.15                &0.05             &0.05                         \\
  $D(1^3D_3)^+\pi^-$    & 0.04               &$-$            &0.27                &0.09             &0.10                         \\
  $D_2(1D)^0\pi^0$        & 0.32               &0.02           &0.51                &1.59             &$5.6\times 10^{-5}$                         \\
  $D_2(1D)^+\pi^-$        & 0.61               &0.03           &1.00                &3.17             &$1.2\times 10^{-4}$                         \\
$D^\prime_2(1D)^0\pi^0$   & 0.06               &1.49           &0.09                &37.24            &$5.9\times 10^{-3}$                         \\
$D^\prime_2(1D)^+\pi^-$   & 0.11               &2.82           &0.17                &74.35            &0.01                         \\
  Total width           & 157.69             & 639.26        &218.76              & 342.87          & 103.90 \\
  Experiment            &\multicolumn{5}{c} {$110.5\pm11.5$}\\
\hline\hline
\end{tabular}
\end{center}
\end{table}

\section{Summary}

In order to understand the possible quark-model assignments of the
BaBar and LHCb charmed states, we apply the GI quark model to
calculate the $1S$, $2S$, $3S$, $1P$, $2P$, $3P$, $1D$, $2D$, $3D$,
$1F$, $2F$, $3F$, $1G$, and $2G$-wave charmed meson spectroscopy.
Our mass predictions, together with the expectations of some other
relativistic quark models, present the mass ranges of the $D$
mesons. From these predicted mass ranges and the parity information,
we tentatively give the possible quark-model assignments for the
BaBar and LHCb charmed states.

To clarify these possible assignments,
we then evaluate the
strong decay behaviors of these charmed states in the framework of the $^3P_0$ model,
where the GI quark model wave functions are employed. In our calculations, the only one free parameter
$\gamma$, the quark pair creation strength, is obtained by fitting to 34 decay modes with specific branching
ratios.

We calculate the two-body strong decay properties of the $2S$, $3S$,
$2P$, $1D$, and $1F$-wave charmed mesons. Comparison of the strong
decay predictions and the measured decay properties, we find that
the observed decay properties of these charmed states can be
reasonably explained. Therefore, we tend to conclude that the
$D(2550)$/$D_J(2580)$, $D^*(2600)$/$D^*_J(2650)$,
$D_J(2740)/D(2750)$, $D^*(2760)$/$D^*_J(2760)$, $D_J(3000)$, and
$D^*_J(3000)$ can be identified as the $D(2^1S_0)$, $D(2^3S_1)$,
$D^\prime_2(1D)$, $D(1^3D_3)$, $D(3^1S_0)$, and $D(1^3F_4)$,
respectively. Further experimental information on the spin-parity
and branching ratios of these charmed states will provide a useful
consistency check for our present assignments.

Our predictions on the masses and strong decays for the $D(3^3S_1)$, $D_2(1D)$, $D(1^3D_1)$, $D(2^3P_0)$, $D(2^3P_2)$, $D_1(2P)$,
$D^\prime_1(2P)$, $D(1^3F_2)$, $D_3(1F)$, and $D^\prime_3(1F)$ will be useful to search for the corresponding
charmed mesons experimentally.

\bigskip
\noindent
\begin{center}
{\bf ACKNOWLEDGEMENTS}\\
\end{center}

This work is partly supported by the National Natural
Science Foundation of China under Grant No. 11105126.

\end{document}